\documentclass[aps,prd,onecolumn,groupedaddress,showpacs,nofootinbib,amssymb]{revtex4}
\usepackage[dvips]{graphicx}
\usepackage{amssymb}
\usepackage{amsmath}
\usepackage{graphicx,,color}
\usepackage{amsfonts}
\usepackage{bm}
\usepackage{cancel}
\usepackage{comment}
\usepackage{mathtools}

\newcommand\be{\begin{equation}}
\newcommand\ee{\end{equation}}

\newcommand\e{\mathrm{e}}
\newcommand\abs[1]{\left| #1 \right|}

\allowdisplaybreaks[4]

\begin{document}

\tolerance=5000

\title{Statistical System based on $p$-adic {numbers}}
\author{Mikoto~Terasawa$^1$\thanks{terasawa.mikoto@b.mbox.nagoya-u.ac.jp} 
and
Shin'ichi~Nojiri$^{1,2}$\thanks{nojiri@gravity.phys.nagoya-u.ac.jp}
}
\affiliation{$^{1)}$ Department of Physics, Nagoya University,
Nagoya 464-8602, Japan \\
$^{2)}$ Kobayashi-Maskawa Institute for the Origin of Particles and
the Universe, Nagoya University, Nagoya 464-8602, Japan %\\
%$^{6)}$ Laboratory for Theoretical Cosmology, Tomsk State University of Control Systemsand Radioelectronics, 634050 Tomsk, Russia (TUSUR)}
}

\tolerance=5000

\begin{abstract}

We propose statistical systems based on $p$-adic numbers. 
In the systems, the Hamiltonian is a standard real number which is given by a map from the $p$-adic numbers. 
Therefore we can introduce the temperature as a real number and calculate the thermodynamical quantities like free energy, 
thermodynamical energy, entropy, specific heat, etc. 
Although we consider a very simple system, which corresponds to a free particle moving in one dimensional space, 
we find that there appear the behaviors like phase transition in the system. 
Usually in order that a phase transition occurs, we need a system with an infinite number of degrees 
of freedom but in the system where the dynamical variable is given by $p$-adic number, even if {degree of freedom} is unity, 
there might occur the phase transition. 
\end{abstract}

%PACS numbers: 04.50.Kd, 95.36.+x, 98.80.-k, 98.80.Cq
%\pacs{04.50.Kd, 95.36.+x, 98.80.-k, 98.80.Cq,11.25.-w}

\maketitle

\section{Introduction}

Real numbers are obtained from rational numbers by the procedure of completion. 
For the completion, we need to define a distance between two numbers, which is the 
absolute value of the difference between two numbers. 
It is possible to define ``absolute value'' in a way different from the definition of the 
absolute value which we use when we define the real number. 
The $p$-adic numbers are obtained by the completion using the $p$-adic absolute 
value \cite{Hensel:1897}, where $p$ is a prime number. 
For a review, see \cite{Brekke:1993gf} and for a recent developments, \cite{Dragovich:2017kge}. 

Let $\mathbb{Q}_p$ be set of all two-sided sequences, $\dots a_2a_1a_0.a_{-1}a_{-2}\dots$, 
where ``$.$'' is a radix point and $a_i\in\mathbb{F}_p\equiv \mathbb{Z}/p\mathbb{Z}$ 
for each $i$, that is, $a_i\in \left\{0,1,2,\ldots,p-1 \right\}$. 
{An element of $\mathbb{Q}_p$ is
\begin{equation}
\label{p0}
x=\cdots a_2a_1a_0.a_{-1}a_{-2}\cdots 
= \cdots a_2 p^2 + a_1 p + a_0 + a_{-1} p^{-1} + a_{-2} p^{-2} \cdots \, ,
\end{equation}
where all but a finite set of digits with negative indices are zero. 
W}e define the \textit{order} $v_p(x)$ of $x$ and 
the \textit{absolute value} (\textit{valuation}) $\abs{x}_p$ as follows, 
\begin{equation}
\label{p1}
v_p(x)\equiv 
\begin{cases}
\infty & \text{$a_i=0$ for all $i$}, \\
\min\{s:a_s\ne 0\} & \text{otherwise};
\end{cases} \, , \quad 
\abs{x}_p\equiv p^{-v_p(x)} \, .
\end{equation}
For example, we find $\ldots,\abs{\dfrac19}_3=9$, $\abs{\dfrac13}_3=3$, $\abs{1}_3=1$, 
$\abs{3}_3=\frac{1}{3}$, $\abs{9}_3=\frac{1}{9}$, $\abs{27}_3=\frac{1}{27}$, etc. 
For the sequence of numbers $\left\{p^n \right\}$, we obtain $\abs{p^n}_p = p^{-n} \to 0$ 
when $n\to +\infty$ and therefore the sequence converges to vanish. 
Then as an example, we find the following expansion by using $\abs{\cdot}_3$, 
\begin{equation}
\label{p2}
 -\frac{1}{2}=\cdots+3^2+3+1 = \sum_{k=0}^{+\infty} 3^k \, , 
\end{equation}
which corresponds to the formal expansion 
$-\frac{1}{2} = \frac{1}{1-3}=1 + 3 + 3^2 + \cdots$. 

The $p$-adic numbers attracted the attentions of the string physicist due to the 
$p$-adic like structure of the string amplitude 
\cite{Freund:1987kt,Freund:1987ck,Frampton:1988kr}. 
After that, the quantum mechanics including the path integral formulation 
and statistical system have been studied 
\cite{Freund:1987ks,Dragovich:2010kj,Djordjevic:1997xn,Djordjevic:2000zy}. 
In these formulations, the path-integrand or the Hamiltonian is also a $p$-adic numbers. 
In the standard statistical physics, the classical Hamiltonian is a standard real number 
or $c$-number and even for the quantum Hamiltonian, we consider the sum over the 
eigenvalues of the Hamiltonian. 
In this sense, the value of the Hamiltonian is a real number. 
The Hamiltonian of the Ising model can be regarded as a map from $\mathbb{Z}_2$ to real 
numbers and the Hamiltonian for the fermionic fields can be a map from the anti-commuting 
Grassmann numbers to real numbers. 
This motivates us to consider the Hamiltonian which is given by a map from $p$-adic 
numbers to real numbers, that is, we consider a system where the dynamical variables are 
$p$-adic numbers but the Hamiltonian is given by real numbers. 
Then we can introduce the temperature $T$ or several coupling constants as 
$c$-numbers and we can investigate the thermodynamical quantities like free nenergy, 
thermodynamical energy, entropy, specific heat, etc.
A natural map from the $p$-adic numbers to real numbers is given by absolute value 
(valuation) in (\ref{p1}). 
Recently in \cite{Sinclair:2020}, a model of the statistical system, where the 
Hamiltonian is given by the distance of two $p$-adic numbers, that is, the absolute value 
of the difference between the two $p$-adic numbers, has been proposed and well-studied. 
The model can be regarded as a $p$-adic analogue of the electrostatics. 
In this paper, we consider the simplest model corresponding to a single free particle 
moving in one dimensional space. 
Although the model is very simple but we show that the model shows rich thermodynamical 
structures and generates phenomena like phase transition in spite that we are considering 
only one degree of freedom. 

In the next section, as a preparation to consider the model, we review on 
the measure of the $p$-adic number in order to define the integration which we 
use to calculate the partition function of the system. 
In Section \ref{SecIII}, we propose the simplest model which corresponds to a free 
particle moving on one-dimensional space and calculate the thermodynamical quantities, 
whose structures are very rich and complicated. 
The calculations in Section~\ref{SecIII} are mainly given numerically. 
In Section~\ref{SecIIIB}, we try to clarify the structure given in Section~\ref{SecIII} 
analytically as possible as we can. 
The last section is devoted to the summary and discussion on the obtained results and 
we speculate some applications. 

\section{Invariant Measure on the Field $\mathbb{Q}_p$ \label{SecII}}

In order to define the integration with respect to the $p$-adic numbers, 
we first consider the invariant measure on $\mathbb{Q}_p$. 
For details, see \cite{Brekke:1993gf}. 

Let assume $a,b \in\mathbb{Q}_p$. 
Then there exists the Haar measure, which is positive and satisfies the conditions, 
\begin{equation}
\label{p3}
d(x+a)=dx\, , \quad d(xb)=\abs{b}_pd {x}\, .
\end{equation}
We normalize this measure so that 
\begin{equation}
\label{p4}
\int_{B_0}dx=1\, .
\end{equation}
{
Here $B_0=B_{\gamma=0}(a=0)$ is a region inside a circle on the $p$-adic number, 
which is defined by, for general $\gamma$ and $a$, 
\begin{equation}
\label{p5}
B_\gamma(a)=\{x:\abs{x-a}_p\le p^\gamma\} \, ,
\end{equation}
and we denote $B_\gamma(a=0)$ by simply $B_\gamma$. }
We also define the circumference of the circle by 
\begin{equation}
\label{p6}
S_\gamma(a)=\{x:\abs{x-a}_p=p^\gamma\}\, .
\end{equation}
A function $f\in L_\text{\rm loc}^1$ is called \textit{integrable} if there exists 
\begin{equation}
\label{p7}
\lim_{N\to\infty}\int_{B_N}f(x)d {x}
=\lim_{N\to\infty}\sum_{-\infty<\gamma\leq N}\int_{S_\gamma}f(x)d {x}\, .
\end{equation}
We also denote the integration by 
\begin{equation}
\label{p8}
\int_{\mathbb{Q}_p}f(x)d {x}=\sum_{-\infty<\gamma<\infty}\int_{S_\gamma}f(x)d {x}\, .
\end{equation}
Then we find the following formula, 
\begin{equation}
\label{p9}
\int_{\mathbb{Q}_p}f(\abs{x}_p)d {x}=\left( 1-\frac{1}{p} \right)
\sum_{-\infty < \gamma<\infty}f(p^\gamma)p^\gamma \, .
\end{equation}
We do not give any proof of the formula (\ref{p9}) but we use this formula to 
obtain the partition functions of the statistical system in the following section. 

\section{A model of a single particle in one dimensional space \label{SecIII}}

We consider the system of {a} single $p$-adic particle, which {corresponds to a} single 
free particle in ideal gas in one dimension, and investigate the 
following partition function in the canonical emsemble, 
\begin{equation}
\label{p10}
Z=\int\frac{d {q}d {x}}{2\pi\hbar}\e^{-\beta H}\, , \qquad H=\abs{q}^2 \, ,
\end{equation}
where $\beta$ is the inverse of the temperature $T$ with the Boltzmann 
constant normalized to be unity and $q$ can be identified 
with the momentum of the particle. 
Then by using the formula (\ref{p9}), we obtain 
\begin{equation}
\label{p11}
Z\propto Z_p \left(\beta\right) 
\equiv \int_{\mathbb{Q}_p}d {q}\, \e^{-\beta\abs{q}_p^2}
=\left(1-\frac{1}{p} \right)
\sum_{-\infty<\gamma<\infty}p^\gamma \e^{-\beta p^{2\gamma}} \, .
\end{equation}
Just for the comparison, we may consider a function where the sum 
$\sum_{-\infty<\gamma<\infty}\cdots $ is replaced by the integration 
$\int_{-\infty}^\infty d {x}\cdots $, 
\begin{equation}
\label{p12}
Z_p^c \left(\beta\right) 
\equiv \left(1-\frac{1}{p} \right) \int_{-\infty}^\infty d {x}p^x \e^{-\beta p^{2x}}
= \left(1-\frac{1}{p} \right) \frac1{\ln p^2} Z_\infty \left( \beta \right) \, , \quad 
Z_\infty \left( \beta \right) \equiv \sqrt{\frac\pi\beta} \, .
\end{equation}
Here $Z_\infty \left( \beta \right)$ is the partition function of the usual single 
free particle moving in one dimensional space. 
We should note that the factor $\left(1-\frac{1}{p} \right) \frac1{\ln p^2}$ 
does not depend on $\beta$ and therefore the thermodynamical energy and the 
specific heat etc. corresponding to $Z_p^c \left(\beta\right)$ do not depend on $p$. 
We should also note that the expression of $Z_p \left(\beta\right)$ has a 
quasi-periodicity as follows, 
\begin{equation}
\label{p13}
Z_p \left(\beta p^2 \right) = p^{-1} Z_p \left(\beta\right) \, .
\end{equation}

\begin{center}

\begin{figure}%[h]
%\centering
\includegraphics[keepaspectratio, scale=0.7, bb=10 10 400 250%450
%9 9 358 434
]{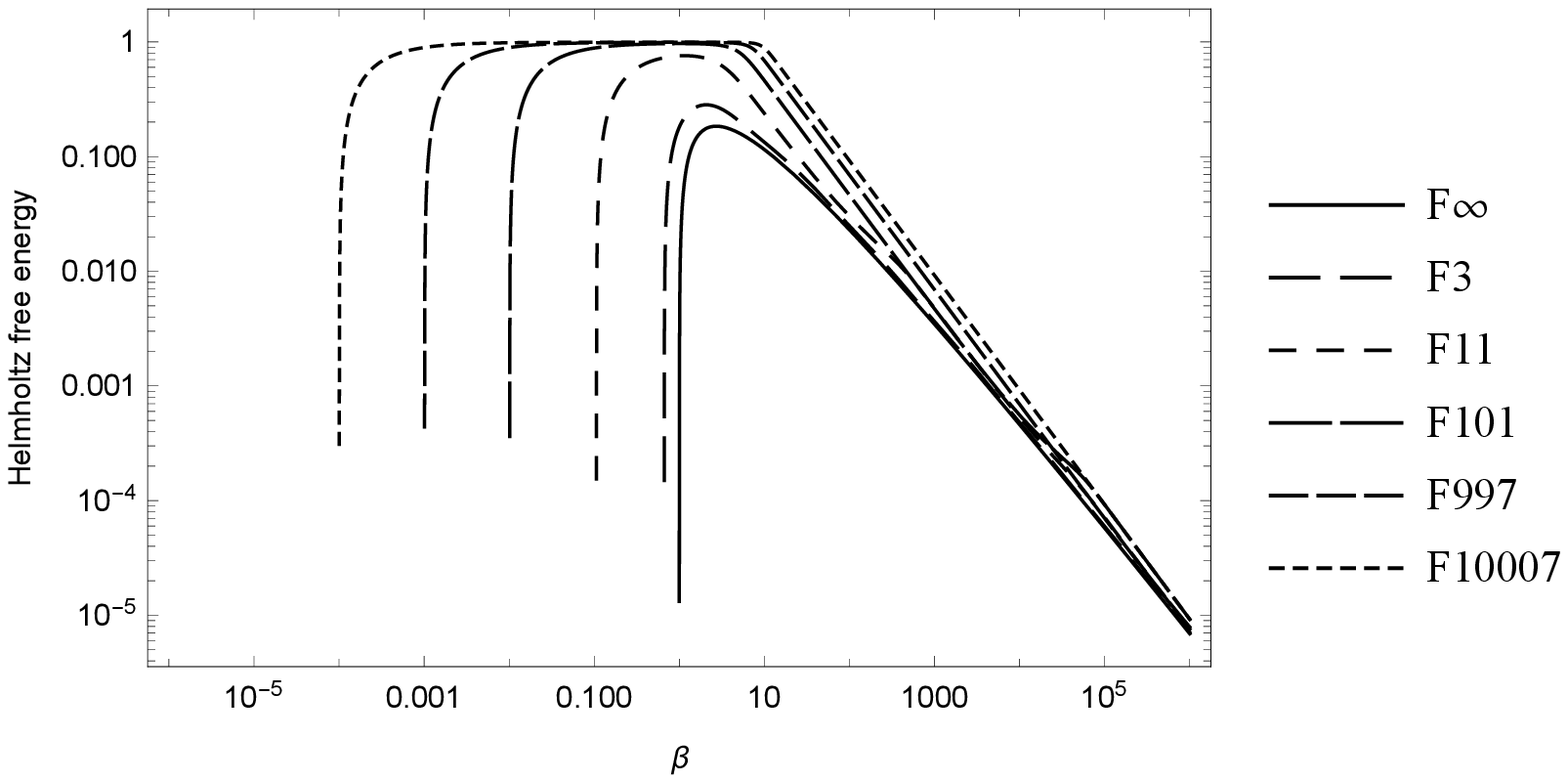}
\caption{
The figure exresses the behaviors of the Helmholz free energies $F_p(\beta)$ for 
$p=3$ $\left(\mathrm{F}3\right)$, $11$ $\left(\mathrm{F}11\right)$, 
$101$ $\left(\mathrm{F}101\right)$, $997$ $\left(\mathrm{F}997\right)$, and 
$10007$ $\left(\mathrm{F}10007\right)$ and $\mathrm{F}\infty$ corresponds to 
$F_\infty =- \beta^{-1} \ln Z_\infty \left(\beta\right)$. 
The vertical axis corresponds to the values of the free energies 
and the horizontal axis to $\beta$. 
}
\label{Fig1}
\end{figure}

\end{center}

The Helmholz free energy is defind by $F_p(\beta) =- \beta^{-1} \ln Z_p \left(\beta\right)$. 
In FIG.~\ref{Fig1}, the free energies for $p=3$, $11$, $101$, $997$, and $10007$ are depicted 
as a function of $\beta$. 
The line for $\mathrm{F}\infty$ corresponds to the free energy defined by 
$F_\infty =- \beta^{-1} \ln Z_\infty \left(\beta\right)$, which is 
nothing but the free energy of the standard (real number) free particle moving 
in one dimensional space. 
The free energies look smooth function of $\beta$ and the difference from $F_\infty$ 
becomes larger if $p$ becomes larger. 

\begin{center}

\begin{figure}%[h]
%\centering
\includegraphics[keepaspectratio, scale=0.7, bb=10 10 400 250
%9 9 358 434
]
{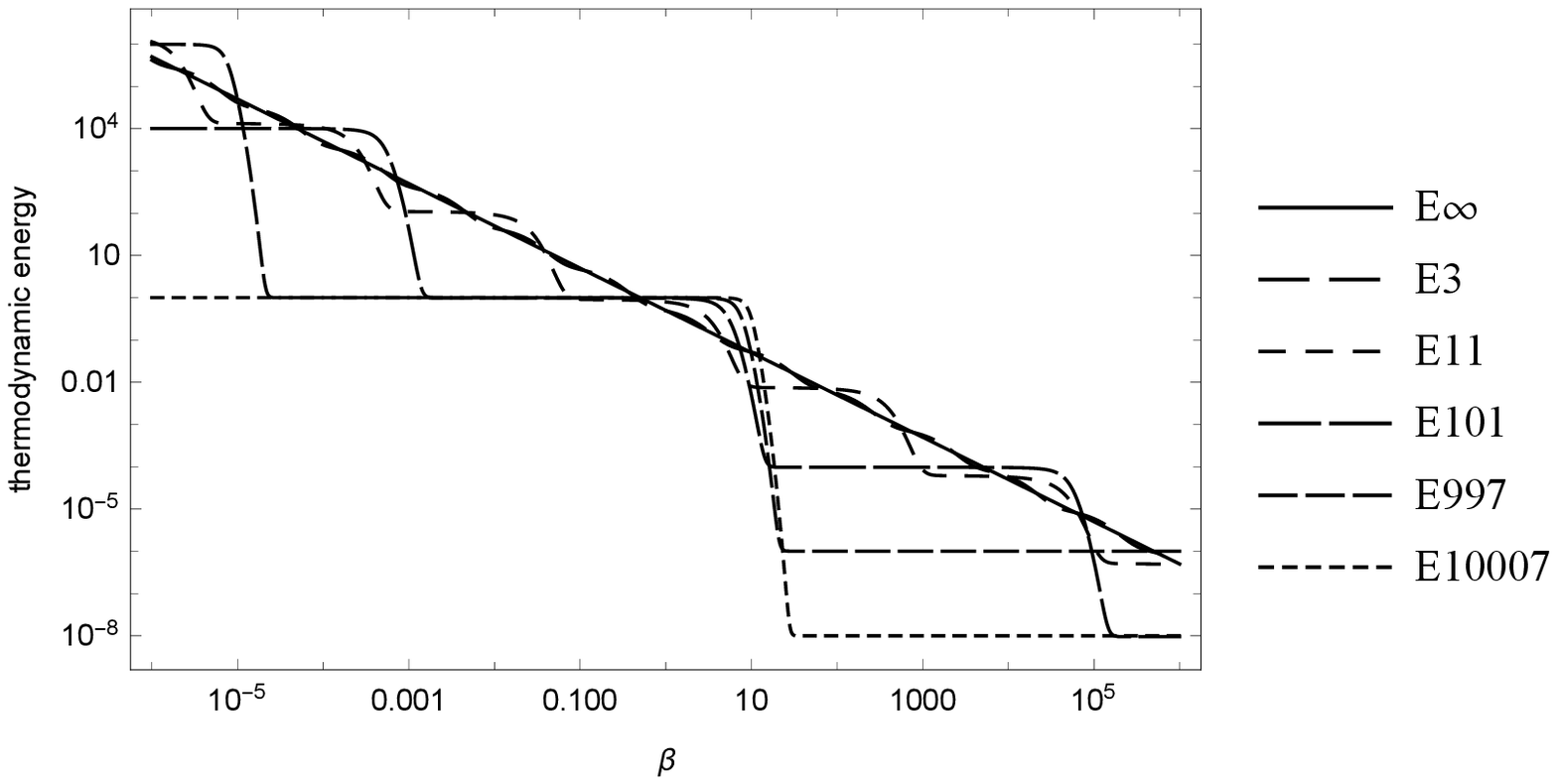}
\caption{
The behaviors of the thermodynamical energies $E_p(\beta)$ for 
$p=3$ $\left(\mathrm{E}3\right)$, $11$ $\left(\mathrm{E}11\right)$, 
$101$ $\left(\mathrm{E}101\right)$, $997$ $\left(\mathrm{E}997\right)$, and 
$10007$ $\left(\mathrm{E}10007\right)$ and $E_\infty$ $\left(\mathrm{E}\infty\right)$ 
are depicted. 
The vertical axis corresponds to the values of the thermodynamical energies 
and the horizontal axis to $\beta$. 
}
\label{Fig2}
\end{figure}

\end{center}

\begin{center}

\begin{figure}%[h]
%\centering
\includegraphics[keepaspectratio, scale=0.7, bb=10 10 400 250
%9 9 358 434
]
{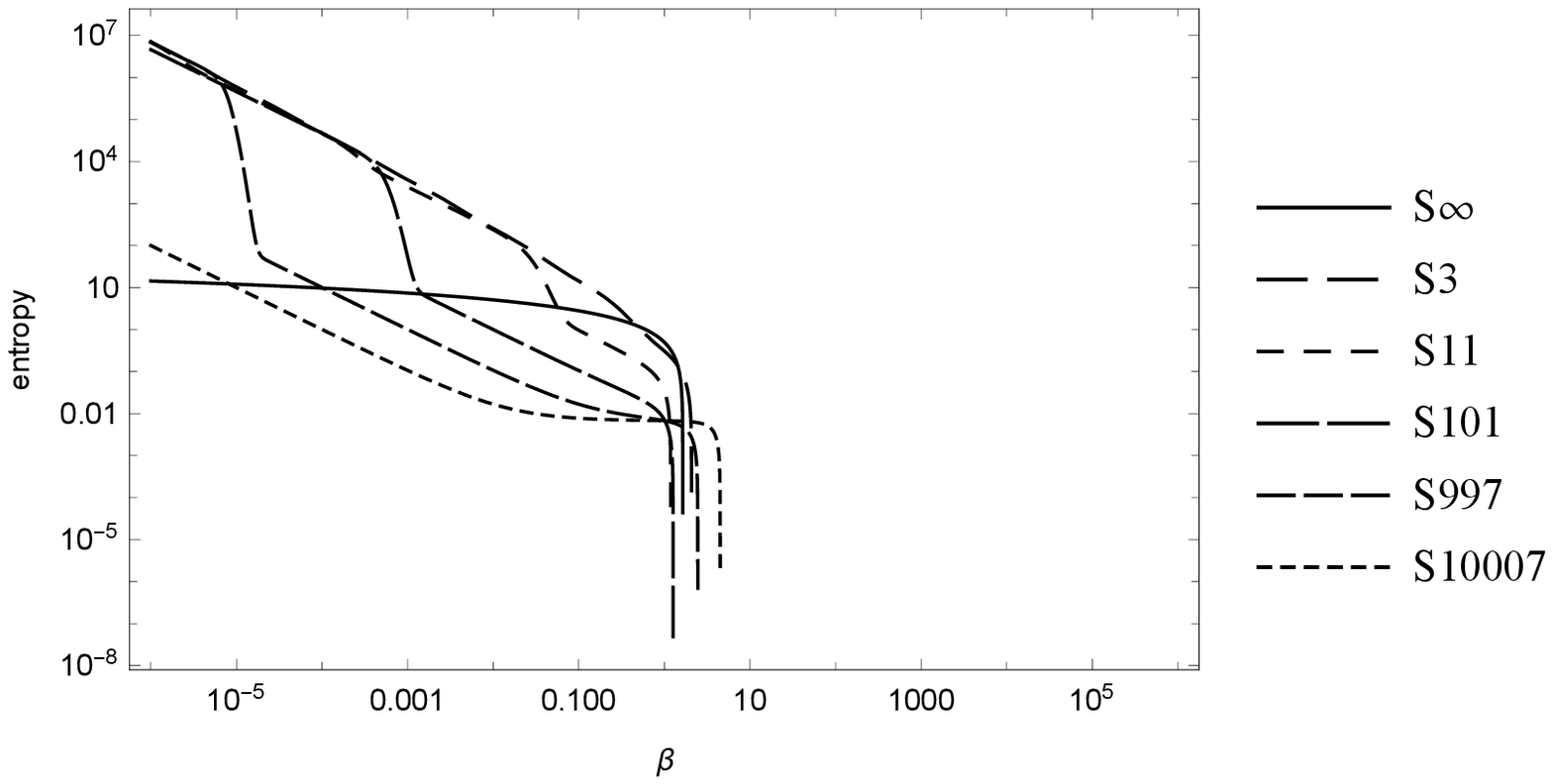}
\caption{
The behaviors of the entropies $S_p(\beta)$ for 
$p=3$ $\left(\mathrm{S}3\right)$, $11$ $\left(\mathrm{S}11\right)$, 
$101$ $\left(\mathrm{S}101\right)$, $997$ $\left(\mathrm{S}997\right)$, and 
$10007$ $\left(\mathrm{S}10007\right)$ and $S_\infty$ $\left(\mathrm{S}\infty\right)$ 
are depicted. 
The vertical axis corresponds to the values of the entropies 
and the horizontal axis to $\beta$. 
}
\label{Fig3}
\end{figure}

\end{center}

\begin{center}

\begin{figure}%[h]
%\centering
\includegraphics[keepaspectratio, scale=0.7, bb=10 10 400 250
%9 9 358 434
]
{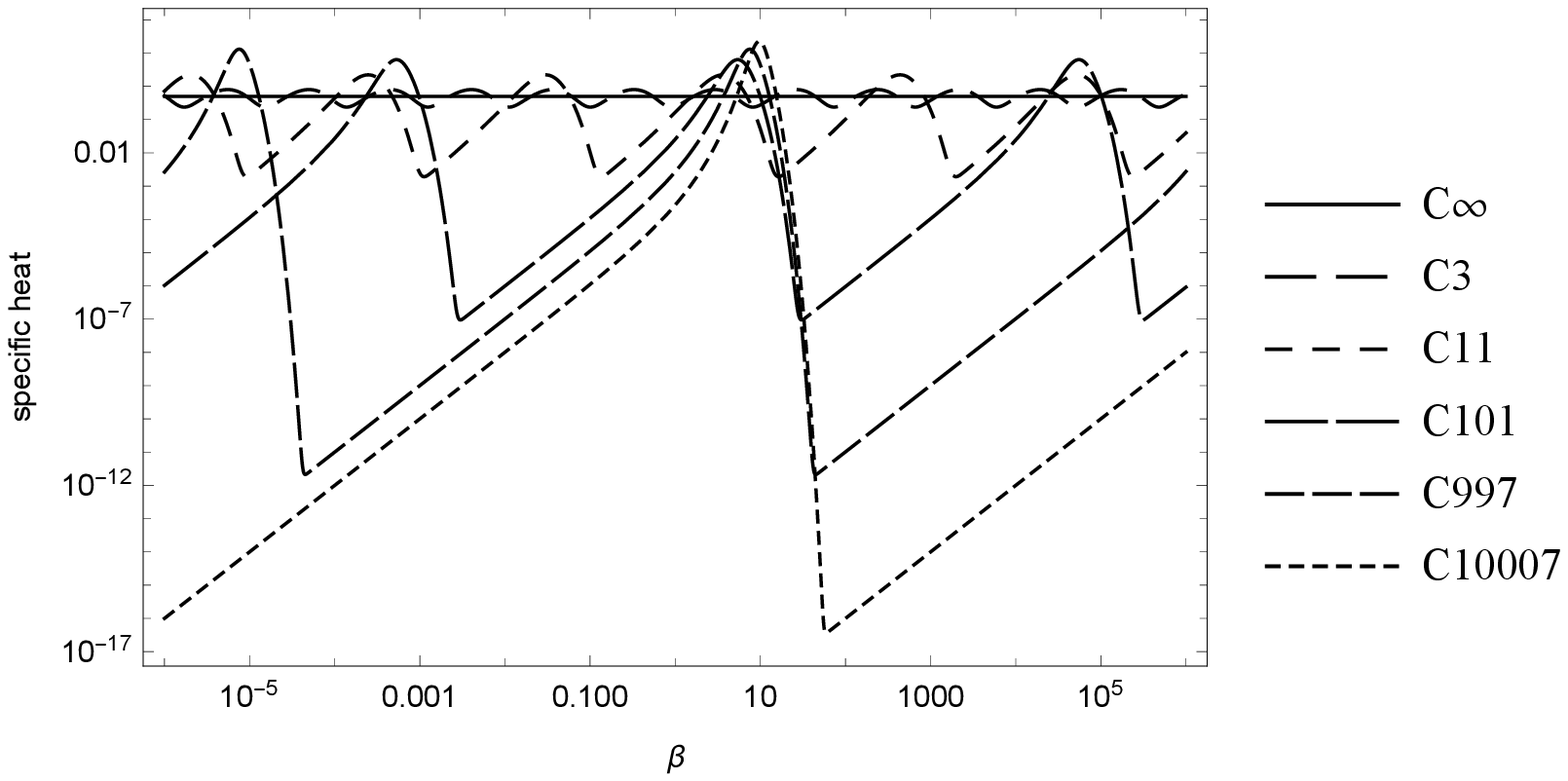}
\caption{
The behaviors of the specific heat $C_p(\beta)$ for 
$p=3$ $\left(\mathrm{C}3\right)$, $11$ $\left(\mathrm{C}11\right)$, 
$101$ $\left(\mathrm{C}101\right)$, $997$ $\left(\mathrm{C}997\right)$, and 
$10007$ $\left(\mathrm{C}10007\right)$ and $C_\infty$ $\left(\mathrm{C}\infty\right)$ 
are depicted. 
The vertical axis corresponds to the values of the specific heats 
and the horizontal axis to $\beta$. 
}
\label{Fig4}
\end{figure}

\end{center}

We may also investigate the thermodynamical energy 
$E_p (\beta) = - \frac{\partial \ln Z_p \left(\beta\right)}{\partial \beta}$, 
the entropy $S_p (\beta) = \beta \left( E_p (\beta) - F_p (\beta) \right)$, 
and the specific heat 
$C_p (\beta) = - \beta^2 \frac{\partial E_p \left(\beta\right)}{\partial \beta}$ and compare 
them with the quantities corresponding to the free particle moving in one dimensional space, 
that is, thermodynamical energy 
$E_\infty (\beta) = - \frac{\partial \ln Z_\infty \left(\beta\right)}{\partial \beta}$, 
the entropy $S_\infty (\beta) = \beta \left( E_\infty (\beta) - F_\infty (\beta) \right)$, 
and the specific heat 
$C_\infty (\beta) = - \beta^2 \frac{\partial E_\infty \left(\beta\right)}{\partial \beta}$. 
In FIG.~\ref{Fig2}, FIG.~\ref{Fig3}, and FIG.~\ref{Fig4}, the behaviors of the thermodynamical 
energies, entropies, and specific heats are depicted, respectively, for $p=3$, $11$, $101$, $997$, and $10007$ as a function of $\beta$.

Although $F_p \left(\beta\right)$ looks a smooth function but 
the thermodynamical energy $E_p (\beta)$, the entropy $S_p (\beta)$, 
and the specific heat $C_p (\beta)$ look to show the oscillation. 
The oscillation could correspond to the quasi-periodicity in (\ref{p13}). 
An interesting point is that there seem to be jumps in the value of 
$E_p (\beta)$, $S_p (\beta)$, and $C_p (\beta)$. 
Because the thermodynamical energy $E_p (\beta)$ is the first derivative of the free 
energy $F_p \left(\beta\right)$, the jumps seem to correspond to the 
first order phase transition. 
Usually{,} in the system with a {finite number of degrees} of freedom, 
phase transitions cannot be generated.

\section{Analytical Properties of Model \label{SecIIIB}}

In the last section, we have found several specific structures for 
the thermodynamical quantities by the numerical calculations. 
In this section, we analyze the behaviors analytically as possible as we can. 

Naively{,} the limit $p\to \infty$ is expected to correspond to the standard 
real number but the results obtained in this paper seem to conflict {with} this 
naive speculation. 
In fact, in any thermodynamical quantity $F_p (\beta)$, $E_p (\beta)$, 
$S_p (\beta)$, or $C_p (\beta)$ which we {have calculated}, 
the difference of the quantity from that in the system of a real free particle, 
$F_\infty (\beta)$, $E_\infty (\beta)$, $S_\infty (\beta)$, or $C_\infty (\beta)$ 
becomes larger when $p$ becomes larger. 
The breakdown of this naive speculation could come from the definition of 
the {absolute value (valuation)} in (\ref{p1}), 
if we fix a value of $q$, the definition (\ref{p1}) tells 
$\left| q \right|_p \to 1$ if $v_p(q)>0$, which tells that 
the region of $q$ which gives a non-trivial contribution to the 
partition function is rather restricted. 
In order to find what happens, we rewrite the r.h.s. {in (\ref{p11})} as below,
\begin{equation}
\label{p11B}
\left(1-\frac{1}{p} \right)
\sum_{-\infty<\gamma<\infty}p^\gamma \e^{-\beta p^{2\gamma}} 
= \left(1-\frac{1}{p} \right)
\sum_{-\infty<\gamma<\infty} \e^{-\beta \e^{2s} + s } \, , 
\quad s\equiv \gamma\ln p
\, .
\end{equation}
If we like to consider the integration corresponding to the free particle in 
one dimensional space as in (\ref{p12}), we need to consider the limit where $ds 
= d\gamma \ln p$ vanishes for a finite $d\gamma$. 
The limit is not the limit of $p\to \infty$ but $\ln p\to 0$, that is, the 
limit of $p\to 1$. 
In the limit, we can replace $\sum_{-\infty<\gamma<\infty} \cdots$ 
by $\frac{1}{\ln p}\int_{-\infty}^\infty ds \cdots$ and we obtain the result in (\ref{p12}). 

Now we consider why the jumps observed in this paper could occur. 
First we estimate which $\gamma$ contributes to the thermodynamical quantities 
by investigating the saddle point in the expression of (\ref{p11B}). 
The saddle point $s=s_0$ is given by 
\begin{equation}
\label{p11C}
0 = \left. \frac{d}{ds} \left( -\beta \e^{2s} + s \right) \right|_{s=s_0}
= -2 \beta \e^{2s_0} + 1 \, ,
\end{equation}
that is, 
\begin{equation}
\label{p11C2}
s_0 = - \frac{1}{2} \ln \left( 2\beta \right) \, .
\end{equation}
Then we find that $s_0$ is monotonically decreasing function of $\beta$. 
We should note, however, that $\gamma_0 \equiv \frac{s_0}{\ln p}$ is not always 
an integer. 
Therefore especially for large $p$, only one of the integer value $\gamma=\gamma_0$ 
which satisfies $\left| \gamma_0 - \frac{s_0}{\ln p}\right| < 1$ dominates and 
we can {use the following approximation,} 
\begin{equation}
\label{p11D}
\sum_{-\infty<\gamma<\infty} \e^{-\beta \e^{2s} + s} 
\sim \e^{-\beta \e^{2\gamma_0 \ln p} + \gamma_0 \ln p} \, .
\end{equation}
If the value of $\beta$ increases, that is, the temperature decreases, 
and goes beyond a critical value, 
the contribution coming from $\gamma=\gamma_0 -1$ becomes larger than that coming from 
$\gamma=\gamma_0$. 
Therefore there occurs a jump in the dominant contribution, which also generates 
the jumps in the thermodynamical quantities. 

For example, when $\beta=\frac{1}{2}$, we find $s_0=0$ and therefore $\gamma=0$. 
\begin{equation}
\label{p11E}
\left. p^\gamma \e^{-\beta p^{2\gamma}} \right|_{\beta=\frac{1}{2}, \gamma=0}=1 \, , \quad 
\left. p^\gamma \e^{-\beta p^{2\gamma}} \right|_{\beta=\frac{1}{2}, \gamma=-1}
=\frac{\e^{-\frac{1}{2p^2}}}{p} \, , \quad 
\left. p^\gamma \e^{-\beta p^{2\gamma}} \right|_{\beta=\frac{1}{2}, \gamma=1}
=p \e^{-\frac{p^2}{2}} \, .
\end{equation}
In the limit of $p\to \infty$, we find 
$\left. p^\gamma \e^{-\beta p^{2\gamma}} \right|_{\beta=\frac{1}{2}, \gamma=-1}$, 
$\left. p^\gamma \e^{-\beta p^{2\gamma}} \right|_{\beta=\frac{1}{2}, \gamma=1} \to 0$ and therefore 
only $\left. p^\gamma \e^{-\beta p^{2\gamma}} \right|_{\beta=\frac{1}{2}, \gamma=0}$ 
contribute. 
This tells that when $\beta\sim \frac{1}{2}$, only the term with $\gamma=0$ dominates 
when $p$ is large. 
On the other hand, when $\beta= \frac{p^2}{2}>\frac{1}{2}$, the term 
with $\gamma=-1$ dominates and we find 
\begin{equation}
\label{p11F}
\left. p^\gamma \e^{-\beta p^{2\gamma}} \right|_{\beta=\frac{p^2}{2}, \gamma=-1}=\frac{\e^{-\frac{1}{2}}}{p} \, , \quad 
\left. p^\gamma \e^{-\beta p^{2\gamma}} \right|_{\beta=\frac{p^2}{2}, \gamma=-2}
=\frac{\e^{-\frac{1}{2p^2}}}{p^2} \, , \quad 
\left. p^\gamma \e^{-\beta p^{2\gamma}} \right|_{\beta=\frac{p^2}{2}, \gamma=0}
= \e^{-\frac{p^2}{2}} \, .
\end{equation}
Therefore in the limit of $p\to \infty$, we find the term 
$\left. p^\gamma \e^{-\beta p^{2\gamma}} \right|_{\beta=\frac{p^2}{2}, \gamma=-1}$ 
dominates. 
If the value of $\beta$ changes from $\beta=\frac{1}{2}$ to $\beta= \frac{p^2}{2}$, there should 
occur a transition where 
{
the dominant contribution changes from the term with $\gamma=0$ to that with $\gamma=-1$.} 
The critical value $\beta_c$, $\frac{1}{2}<\beta_c<\frac{p^2}{2}$, is given by solving the equation 
\begin{equation}
\label{p11G}
\left. p^\gamma \e^{-\beta_c p^{2\gamma}} \right|_{\gamma=0}
=\left. p^\gamma \e^{-\beta_c p^{2\gamma}} \right|_{\gamma=-1} \, ,
\end{equation}
that is, 
\begin{equation}
\label{p11GG}
\e^{-\beta_c} =\frac{\e^{-\frac{\beta_c}{p^2}}}{p} \, ,
\end{equation}
whose solution is given by 
\begin{equation}
\label{p11H}
\beta_c = \frac{\ln p}{1 - \frac{1}{p^2}} \, .
\end{equation}
Therefore we obtain $\beta_c\sim 5$ for $p=997$ and $\beta_c \sim 10$ for $p=10007$, 
which may correspond to the behaviors arround $\beta \sim10$ in FIGs.~\ref{Fig2} and \ref{Fig4}. 
The generalization of the critical value $\beta_c$ corresponding to the transition between 
$\gamma=\gamma_0$ and $\gamma=\gamma_0 - 1$ can be obtained by solving the equation 
\begin{equation}
\label{p11I}
p^{\gamma_0} \e^{-\beta_{c\, \gamma_0} p^{2\gamma_0}} 
= p^{\gamma_0 -1} \e^{-\beta_{c\, \gamma_0} p^{2\left(\gamma_0 - 1\right)}} \, ,
\end{equation}
as
\begin{equation}
\label{p11J}
\beta_{c\, \gamma_0} = \frac{\ln p}{p^{2\gamma_0} \left(1 - \frac{1}{p^2}\right)} \, .
\end{equation}
The transition from $\gamma=\gamma_0$ to $\gamma=\gamma_0 -1$ is very similar to the 
standard first order phase transition and therefore 
the expectation value of $\gamma$ could be the order parameter 
{specifying} the phases.

\section{Summary and Discussions \label{SecIV}}

In this paper, we {have investigated} the thermodynamics of the simplest model given in (\ref{p10}), 
where the dynamical variable $q$ is a $p$-adic number but the Hamiltonian 
is given by a real number. 
Although the degree of freedom is unity, the system shows the {behaviors} like 
phase transition and we have found that the system has rich structures. 
Anyway at present, the physical meaning of the jump in the thermodynamical energy is 
still not clear although we have given some analyitical arguments. 
Maybe we need to clarify it in future works for {further} understanding of the models. 

Similar to the situation that the fermion fields are described by the Grassmann 
number, there could be a situation that some fields are described by the $p$-adic numbers. 
Such theories might be realized by considering a lattice instead of the continuous 
space-time as in the lattice field theories, and putting the $p$-adic dynamical 
degrees of freedom on the sites of the lattice. 
If there exists a model which generates the second order phase transition {corresponding to the 
continuum limit}, we may obtain 
the $p$-adic field theory. 

The behavior of the thermodynamical energy might be interesting if we consider the 
cosmology. 
For large $p$, the energy is almost constant in the large range of $\beta$ and when 
$\beta$ becomes large enough, that is, the temperature {becomes} low enough, there appears 
a jump in the value of the energy and the value becomes much smaller. 
The constant energy might play the role of the cosmological constant. 
Then the large constant value of the thermodynamical energy for high temperature 
(small $\beta$) might generate the inflation in the early universe 
and the small constant energy for the 
low temperature (large $\beta$) might correspond to the dark energy in the 
present universe.

\begin{acknowledgments}
This work is partially supported by the JSPS Grant-in-Aid for Scientific Research (C) No. 18K03615
(S.N.).
\end{acknowledgments}

\end{document}